\documentclass[iop]{emulateapj}
\usepackage{amsmath}
\usepackage{epstopdf}
\usepackage{natbib}
\usepackage{apjfonts}

\newcommand{\avg}[1]{{\left\langle{#1}\right\rangle}}
\newcommand{\bootes}{Bo\"{o}tes} 
\newcommand{\xbootes}{XBo\"{o}tes} 
\slugcomment{Accepted for publication in The Astrophysical Journal}
\shorttitle{SMBH accretion rate and SFR Correlation in SF Galaxies}
\shortauthors{CHEN ET AL.}
\begin{document}
\title{A Correlation between star formation rate and average black hole accretion in star forming galaxies}
\author{Chien-Ting J. Chen\altaffilmark{1}}
\author{Ryan C. Hickox\altaffilmark{1}}
\author{Stacey Alberts\altaffilmark{2}}
\author{Mark Brodwin\altaffilmark{3}}
\author{Christine Jones\altaffilmark{4}}
\author{Stephen S. Murray\altaffilmark{5}}
\author{David M. Alexander \altaffilmark{6}}
\author{Roberto J. Assef\altaffilmark{7,8}}
\author{Michael J.~I. Brown\altaffilmark{9}}
\author{Arjun Dey\altaffilmark{10}}
\author{William R. Forman\altaffilmark{4}}
\author{Varoujan Gorjian\altaffilmark{7}}
\author{Andrew D. Goulding\altaffilmark{4}}
\author{Emeric Le Floc'h\altaffilmark{11}}
\author{Buell T. Jannuzi\altaffilmark{10}}
\author{James R. Mullaney\altaffilmark{6,11}}
\author{Alexandra Pope\altaffilmark{2}}

\altaffiltext{1}{Department of Physics and Astronomy, Dartmouth College, 6127 Wilder Laboratory, Hanover, NH 03755, USA; ctchen@dartmouth.edu.}
\altaffiltext{2}{Department of Astronomy, Amherst, University of Massachusetts, Amherst, MA 01003, USA} 
\altaffiltext{3}{Department of Physics and Astronomy, University of Missouri, 5110 Rockhill Road, Kansas City, MO 64110}
\altaffiltext{4}{Harvard-Smithsonian Center for Astrophysics, 60 Garden Street,
 Cambridge, MA 02138.}
\altaffiltext{5}{Department of Physics \& Astronomy, The Johns Hopkins University, 3400 N.\ Charles Street, Baltimore, MD 21218.}
\altaffiltext{6}{Department of Physics, Durham University, South Road, Durham, DH1 3LE, United Kingdom}
\altaffiltext{7}{Jet Propulsion Laboratory, California Institute of Technology, Pasadena, CA 91109.}
\altaffiltext{8}{NASA Postdoctoral Program Fellow.}
\altaffiltext{9}{School of Physics, Monash University, Clayton 3800, Victoria, Australia.}
\altaffiltext{10}{National Optical Astronomy Observatory, Tucson, AZ  85726.}
\altaffiltext{11}{Laboratoire AIM-Paris-Saclay, CEA/DSM/Irfu - CNRS - Universit\'{e} Paris Diderot, CE-Saclay, pt courrier 131, 91191 Gif-sur-Yvette, France}

\begin{abstract}
We present a measurement of the average supermassive black hole accretion rate (BHAR) as a function of star formation rate (SFR) for galaxies in the redshift range $0.25<z<0.8$. We study a sample of 1,767 far-IR selected star-forming galaxies in the 9 deg$^2$ \bootes\ multiwavelength survey field. The SFR is estimated using $250\;\micron$ observations from the {\it Herschel Space Observatory}, for which the contribution from the AGN is minimal. In this sample, 121 AGNs are directly identified using X-ray or mid-IR selection criteria. We combined these detected AGNs and an X-ray stacking analysis for undetected sources to study the {\it average} BHAR for all of the star-forming galaxies in our sample.
We find an almost linear relation between the average BHAR (in $M_{\odot}\;\textmd{yr}^{-1}$) and the SFR (in $M_{\odot}\;\textmd{yr}^{-1}$) for galaxies across a wide SFR range $0.85<\log\textmd{SFR}<2.56$ : $\log \textmd{BHAR}=(-3.72\pm0.52)+(1.05\pm0.33)\log\textmd{SFR}$. This global correlation between SFR and average BHAR is consistent with a simple picture in which SFR and AGN activity are tightly linked over galaxy evolution timescales.
\end{abstract}


\section{Introduction}
Observational studies have shown that the mass of galactic bulges is tightly correlated with the mass of their central supermassive black holes (SMBHs) \citep[e.g.][]{mago98,ferr00,gebh00,marc04smbh}, and that the black hole accretion rate (BHAR) density and star formation rate (SFR) density both peak at a  similar redshift before declining to the present day \citep[e.g.][]{hopk06sfhist,rodi11irlf,silv09sfagn,aird10xlf}. Together, these results may imply parallel evolutionary paths for the growth of SMBHs and the stellar mass of their host galaxies. However, the physical mechanisms that drive this apparent link between SF and BH growth over a wide variety of galaxies are still poorly understood. \par

Active SMBH accretion (i.e. active galactic nucleus, AGN) and galactic star formation (SF) both require a supply of gas. Thus, the clues of uncovering the connection between their growth may lie in the gas fueling mechanisms that supply both galactic star formation and AGN. 
Recent studies have observed the existence of two different modes of star formation: the quiescent ``main sequence'' star formation, and starbursts \citep{genz10,elba11ms_aph}. The first mode can be fueled by continuous gas inflow \citep{kawa08,deke09,ciot10flare} and the second mode is postulated to be triggered by gas-rich major mergers \citep[e.g.][]{hopk09fueling,veil09qsomorph}. However, it is still not clear whether 
these processes also drive the growth of SMBHs, or whether SMBH growth would scale similarly with galactic star formation in different SF modes, since the dynamical scales of the gas inflows that induce galactic star formation and SMBH accretion are vastly different \citep[see][for a review]{bhreview12}. \par 

A number of studies have investigated the link between SFR and BHAR. For high-luminosity AGNs, an increase in the average SFR as a function of BHAR has been observed \citep[e.g.][]{lutz08qsosf,serj09qsosf,serj10,mor12}, while other studies have also found weak or inverted connections \citep{page12,harr12}. Studies with inclusions of lower luminosity AGNs further suggest that the evolutionary link between SMBHs and their host galaxies only exists in high luminosity AGNs that are possibly triggered by mergers, and there is little or no correlation at lower luminosities \citep[e.g.][]{shao10agnsf,lutz10agnsf,rosa12agnsf,rovi12}.
On the contrary, the study of the {\it average} BHAR of star-forming galaxies implies that the galaxy and SMBH growth rates may be strongly connected when averaging over the whole population of star-forming galaxies \citep{raff11agnsf_aph, mull12agnsf}. Thus, whether BH growth follows SF in all galaxies, or only in the most powerful systems, remains a matter of debate. \par

The apparent contradictory results may be attributed to the difference in the characteristic timescales of SF and BH accretion. Theoretical studies of SMBH accretion with feedback imply that the SMBH accretion rate can vary by more than five orders of magnitudes on timescale of less than 1 Myr \citep{nova11}. 
Observational studies of AGNs also suggest that the observed AGN Eddington ratios range from $<10^{-4}$ to $\sim 1$ for AGN hosts with similar properties \citep{hickox09,hopk09qso,aird12agnedd,bong12}. 
Recent evidence for rapid AGN variability over a large dynamic range in accretion rate comes from the discovery of giant ionized clouds around galaxies with little or no current AGN activity \citep{scha10morph}, indicating a drop in AGN luminosity by $>10^5$ in $<10^5$ years. \citep{keel12a,keel12b}. There is also observational evidence in the X-ray Fe K$\alpha$ echoes from the quiescent SMBH in the center of the Milky Way, suggesting it might have been a low-luminosity AGN a few hundred years ago \citep[e.g.][]{revn04mky,pont10mky,nobu11mky}. In contrast to the accretion rate of an individual SMBH, the galactic SFR is relatively stable. Even short-lived starbursts last $\sim 100$ Myr \citep[e.g.][]{wong09sfr,ostr10,hick12smg}, which is still much longer than the timescale of AGN variability. \par
The key quantity to study may therefore be the {\em average} AGN luminosity of a population, which thus smoothes over the variations of individual sources. Recent studies \citep[e.g.][]{hopk09qso,aird12agnedd,bong12} have discovered that the shape of the distribution function of the AGN Eddington ratio is independent of the properties of the hosting galaxies, thus the average AGN luminosity is a reliable tool to study the long-term black hole accretion rate in any sample of galaxies. \par

Measuring star formation rates in AGN host galaxies can be challenging because of obscuring dust and contamination from AGN. 
Detailed spectral template fitting is required to disentangle AGN and star forming activities in optical or mid-IR observations \citep[e.g.][]{kauf03host,pope08smgspitz}. In contrast, at far-IR wavelengths, at which the thermal emission from the cold dust peaks, the AGN contaminates the least \citep[e.g.][]{netz07qsosf,hatz10,mull11agnsf}.
Therefore, the $250\;\micron$ filter on the Spectral and Photometric Imaging Receiver \citep[SPIRE,][]{spire} on board the {\it Herschel Space 
Observatory} provides an excellent tool to probe the dust-enshrouded star formation activities of galaxies hosting AGNs at $z<1$.\par

For this paper, we utilized $250\;\micron$ {\it Herschel} SPIRE observations to constrain the SFR of a sample of star-forming galaxies with spectroscopic redshift measurements from the AGN and Galaxy Evolution Survey \citep[AGES,][]{AGES11}. We also supplemented the AGES redshift measurements with photometric redshifts from {\it Spitzer} Deep Wide Field Survey \citep[SDWFS,][]{ashb09sdwfs} and the {\it Spitzer} IRAC (Infrared Array Camera) Shallow Survey \citep[ISS,][]{brod06photoz}. We focused on studying the correlations between the SFR and the BHAR of galaxies in a redshift range of $0.25<z<0.8$.  To determine the connection between the SF activity and the {\it average} SMBH accretion rate, we measure AGN luminosities using a combination of X-ray and mid-IR observations from the {\it Chandra} XBo\"{o}tes \citep{murr05,kent05} survey, and the {\it Spitzer} ISS and SDWFS catalogs . For galaxies without identified AGNs, we employed an X-ray stacking analysis. 
\par

This paper is organized as follows: in \S 2 we describe the multi-wavelength data and the properties of the galaxy and AGN samples, along with the methods we adopted to obtain their SFR and average X-ray luminosity. The results of our SFR and SMBH accretion rate analysis is presented in \S 3, and we provide a discussion and a summary in \S 4. Throughout the paper, we assume a $\Lambda$CDM cosmology with $\Omega_m=0.3$ and $\Omega_\Lambda=0.7$. For direct comparison with other works, we assume $H_0=70$ km s$^{-1}$ Mpc$^{-1}$, however, our conclusions are insensitive to the exact choice of cosmological parameters.

\section{Data and Sample Selection}
\subsection{Redshifts}
We selected {\it Herschel}-observed star-forming galaxies in the 9 deg$^2$ Bo\"{o}tes field covered by the NOAO Deep Wide-Field Survey \citep[NDWFS.][]{jann99} to measure the correlation between the SFR and the BHAR in star-forming galaxies. For the redshifts in this study, we primarily used the spectroscopic redshifts in the  range $0.25<z<0.8$ from the AGN and Galaxy Evolution Survey (AGES) Data Release 2 \citep{AGES11}, which covers 7.7 deg$^2$ of NDWFS. To maximize the completeness for our sample of IR-selected SF galaxies, we also supplemented the data with photometric redshifts measured using the data from the 8.5 deg$^2$ {\it Spitzer} IRAC Shallow Survey \citep[ISS,][]{eise04,ster05} and the 10 deg$^2$ {\it Spitzer} Deep Wide Field Survey \citep[SDWFS,][]{ashb09sdwfs}. The photometric redshifts were derived using all four IRAC bands (3.6, 4.5, 5.8 and 8 $\micron$) in ISS and SDWFS, with the algorithm developed by \citet[B06 hereafter]{brod06photoz}. We limited our photometric redshifts to the same redshift range as the spectroscopic redshifts, in which the accuracy of the photometric redshifts is $\sigma=0.06(1+z)$ for $95\%$ of galaxies and $\sigma=0.12(1+z)$ for $95\%$ of AGNs \citep{brod06photoz}. \par

\subsection{Infrared Data}
In addition to the IRAC observations, we also used far-IR and mid-IR data from {\it Herschel} and the Multiband Imaging Photometer (MIPS) onboard {\it Spitzer}, respectively. The far-IR data in this work is based on the publicly available {\it Herschel} SPIRE $250\;\micron$ observations from the {\it Herschel} Multi-tiered Extragalactic Survey \citep[HerMES,][]{hermes}.  We re-reduced and mosaiced the \bootes\ SPIRE observations (Alberts et al., in preparation), which include a deep $\sim2$ deg$^2$ inner region in the center of the field and a shallower $\sim8.5$ deg$^2$ outer region. We specifically focused on removing stripping, astrometry offsets, and glitches missed by the standard pipeline reduction. We also convolved the raw maps with a matched filter \citep[see][]{chap11}, which aided in source extraction by lowering the overall noise and de-blending sources.  From this, we generated a matched filter catalog with 21,892 point sources above SNR $> 5$.  Completeness simulations showed that these catalogs are $95\%$ complete in the inner region and $69\%$ complete in the outer regions above a flux limit of 20 mJy. We also found minimal flux boosting for low SNR sources above these flux cut-offs. In addition, we used the $24\;\micron$  flux measurements available from the Multiband Imaging Photometer for {\it Spitzer} (MIPS) GTO observations (IRS GTO team, J. Houck (PI), and M. Rieke) of the Bo\"{o}tes field as a comparison in the source matching and SFR estimation. This catalog covers an area similar to that of the \xbootes\ and has 52,089  SNR $>5$ sources with flux $>0.15$ mJy. 

\begin{figure}
\epsscale{1.4}
\plotone{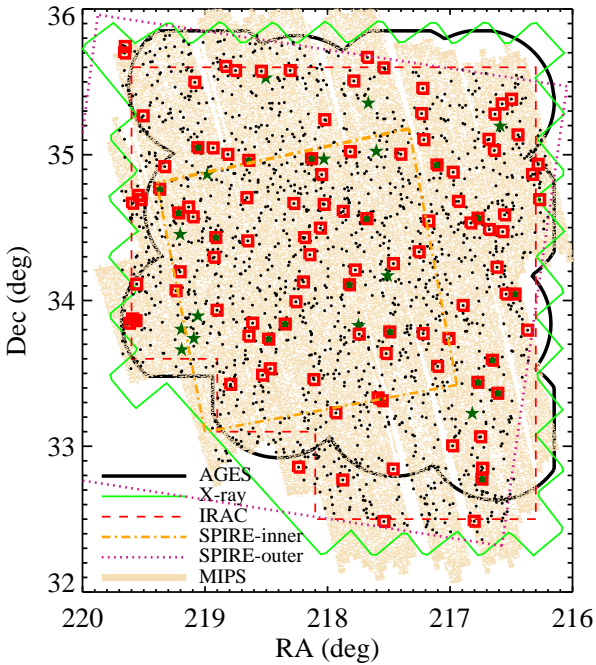}
\caption{Map of the Bo\"{o}tes survey region, showing approximate areas covered by the AGES, XBo\"{o}tes (X-ray) , IRAC Shallow Survey (mid-IR) and the inner deeper and the outer shallower regions of HerMES (far-IR). The orange shaded region is the area covered by MIPS $24\;\micron$ observations. X-ray and mid-IR AGNs are also marked as green stars and red squares in the plot. }
\label{fig:field}
\end{figure}

\subsection{AGN Selection}
X-ray observations from the XBo\"{o}tes survey provide the basis of our measurements of AGN luminosity. The X-ray data in this study is drawn from the 9.3 deg$^2$ XBo\"{o}tes survey, which is a mosaic of 126 short (5 ks) {\it Chandra} ACIS-I images \citep{murr05,kent05} covering the entire NDWFS. XBo\"{o}tes contains 2,724 X-ray point sources with at least four counts in the AGES survey region. 362 of 2,724 X-ray point sources are not close to bright stars and matched within $3.5''$ to objects with good AGES redshifts at $0.25<z<0.8$ \citep{kent05,hickox09}. These X-ray point sources have 0.5-7 keV luminosities of $10^{42}<L_\textmd{X}<10^{45}$ erg s$^{-1}$, which are characteristic for moderate- to high-luminosity AGNs and significantly larger than the typical $L_\textmd{X}$ for star forming galaxies \citep{rana03sf}.\par
To properly estimate the average SMBH accretion rate, it is important to account for obscured AGNs that may have their X-ray flux significantly depressed by photoelectric absorption \citep[e.g.][]{alex08compthick}. We therefore supplemented our AGN sample with AGNs selected with the IRAC color-color selection criteria \citep{ster05} in ISS and SDWFS. These selection criteria has been shown to effectively identify obscured AGN at moderate redshift \citep{gorj08agn, asse10}, since mid-IR wavelengths are not as strongly affected by obscuration as the optical or UV \citep[e.g.][]{lacy04,ster05,donl07plaw,hick07obsagn,goul09irs}. In our sample, 1,047 AGNs were identified using mid-IR observations (mid-IR AGNs hereafter). Among the 1,047 mid-IR AGNs and the 362 X-ray AGNs, 163 of them can be identified as AGN using both X-ray and mid-IR selection criteria. \par

\subsection{Catalog Matching}
Since the large point-spread function of SPIRE can lead to spurious matching results, here we discuss our catalog matching procedures. We first matched the entire SPIRE catalog to the $5\sigma$ MIPS $24\;\micron$ catalog with a matching radius of $5\arcsec$. We found that $\sim 80\%$ of the SPIRE sources in the coverage of the MIPS catalog have MIPS counterparts within the $5\arcsec$ radius. While increasing the radius to $10\arcsec$ can increase the MIPS detection fraction for the SPIRE sources to $\sim 92\%$, the fraction of SPIRE sources with multiple MIPS counterparts would also increase from $2.9\%$ to $14\%$; therefore, we chose the $5\arcsec$ matching radius to avoid spurious matching. These matched sources have a minimum $24\;\micron$ flux at least $25\%$ larger than the  $0.15$ mJy flux limit of the $5\sigma$ MIPS $24\;\micron$ catalog,  which ensures that the matched galaxies are star-forming galaxies bright in both mid-IR and far-IR, and that our completeness in $250\;\micron$ is not strongly affected by the $24\;\micron$ flux limit. Then we matched the coordinates from the MIPS catalog (with $250\;\micron$ counterparts) to the B06 photometric redshift catalog with a matching radius of $2\arcsec$. We also matched the AGES catalog and the AGES-matched optical positions of X-ray AGNs to the photometric redshift catalog with a matching radius of $1\arcsec$. \par
To minimize spurious matches, we tested our matching by offsetting the source positions by $1\arcmin$ in a random direction. We found that with radius of $1\arcsec$, our matching between the AGES and B06 catalog yielded $<0.1\%$ spurious matches. The radius of the matching between the $250\;\micron$ matched MIPS catalog and the B06 catalog is tested to have less than $4\%$ of spurious matches, which greatly reduced the $\sim 25\%$ spurious matching rates obtained when directly matching the SPIRE sources to the B06 catalog. \par
To minimize spurious matches, we tested our matching by offsetting the source positions by $1\arcmin$ in a random direction. We found that with radius of $1\arcsec$, our matching between the AGES and B06 catalog yielded $<0.1\%$ spurious matches. The radius of the matching between the $250\;\micron$ matched MIPS catalog and the B06 catalog is tested to have less than $4\%$ of spurious matches, which greatly reduced the $\sim 25\%$ spurious matching rates obtained when directly matching the SPIRE sources to the B06 catalog. \par
In our sample, there are 1,767 galaxies with both SPIRE and MIPS detections. 1,112 of these have spectroscopic redshift measurements while the remainder have photometric redshifts. 121 of the 1,767 sources ($\sim 7\%$) in our sample have been classified as AGN, in which 34 ($\sim 2\%$) sources are identified as X-ray AGN using X-ray selection criteria and 107 ($\sim 6\%$) are identified as mid-IR AGNs using mid-IR color-color selection criteria. 20 of the 121 AGNs are identified as both X-ray AGN and mid-IR AGN. The angular distribution of sources and the coverage of different observations are plotted in Fig.~\ref{fig:field}.

\subsection{Star Formation Rate}
In this section, we discuss the methods we used to estimate the star formation rates with the SPIRE $250\;\micron$ observations.\par

Extrapolation of total IR luminosities ($L_\textmd{IR}$, defined as the integrated luminosity in the $8-1000\;\micron$ range) from monochromatic fluxes in the near- to mid-IR wavelengths using templates of infrared spectral energy distributions of local star-forming galaxies \citep[SEDs, e.g.][CE01 hereafter]{ce01} have been adopted by a number of previous works to estimate the SFR of galaxies. However, it is also shown that AGNs have significant emission at near- to mid-IR wavelengths \citep[e.g.][]{dadd07sf,mull11agnsf}, which poses challenges in obtaining reliable estimates of the SFR using mid-IR data. \par
By contrast, far-IR observations have been shown to suffer minimum contamination from AGN \citep[e.g.][]{netz07qsosf,hatz10,mull11agnsf,kirk12}, which makes far-IR emission a better tracer of the star formation related $L_\textmd{IR}$ in an AGN-hosting galaxy sample. However, recent studies have discovered that estimates using only one SPIRE band and templates based on local star-forming galaxies can substantially overestimate $L_\textmd{IR}$ for $z<1.5$ \citep{elba10herschel,nord12} due to the possible lower dust temperature \citep[e.g.][]{dh02,pope06} that was not accounted for in the local templates. Therefore, in this work, we adopted the composite SED template from \citet[hereafter K12]{kirk12}. The $z\sim 1$ K12 template is derived from the {\it Spitzer} and {\it Herschel} observations of a sample of star-forming galaxies at $0.4<z<1.47$, which is comparable to the redshift range of our sample. 
From the photometric observations available in both K12 and our sample, we have determined that the distributions of  the ratio between the observed $250\;\micron$ flux ($S_{250}$) and $24\;\micron$ flux ($S_{24}$) in both samples are similar. We found that $\sim 96\%$ of the star-forming galaxies in our sample have $S_{250}/S_{24}$ that lies within the $2\sigma$ range (0.34 dex, adopted from Table 3 in K12) of the K12 template $S_{250}/S_{24}$ distribution derived in the redshift range of our sample.
This shows that the K12 template can describe the far-IR to mid-IR color in our sample well, thus we can use this template to estimate the total $L_\textmd{IR}$ in our sample.\par
In principle, the far-IR part of an SED for star-forming galaxies, which comprises the bulk of the star-formation related $L_\textmd{IR}$, are dominated by the thermal radiation due to cold and warm dust. Thus the ratio between the monochromatic far-IR flux and the total $L_\textmd{IR}$ of the SED should be very similar for star-forming galaxies with similar dust temperatures. In particular, it has also been shown that the star-forming galaxies with and without strong AGNs have similar cold dust temperature \citep{kirk12}. Recent studies using {\it Herschel} observations have also shown that even for AGN host galaxies, the far-IR ($\geq 100\micron$) emissions are still dominated by the cold dust component \citep[e.g.][]{hatz10,kirk12}. 
Thus, we can estimate the SF-related $L_\textmd{IR}$ for our sample by normalizing the total infrared luminosity of this template ($L_\textmd{IR}^\textmd{T}$) using the $250\;\micron$ observations. For each source in our sample, we calculated the ratio between the observed $S_{250}$ and the monochromatic flux of the template at the corresponding observed-frame $250\;\micron$ ($S_{250}^\textmd{T}$), and derived the total $L_\textmd{IR}$ with the following equation:

\begin{equation} \label{eq:lir}
L_\textmd{IR}=\frac{S_{250}}{S_{250}^\textmd{T}}L_\textmd{IR}^\textmd{T}.
\end{equation}
For the K12 template we chose for this work, $L_\textmd{IR}^\textmd{T}=4.2\times 10^{11}L_\odot$, which corresponds closely to the median $L_\textmd{IR}$ of our sample.\par

In this work, the $L_\textmd{IR}$ was derived by assuming that the $S_{250}/L_\textmd{IR}$ in our sample is similar to that of the K12 composite. For a population of star-forming galaxies, the observed dispersion in $S_{250}/L_\textmd{IR}$ depends on the variations in their SED shapes. This dispersion also depends on the redshift since the observed $S_{250}$ traces the  SED at different rest-frame wavelengths. Thus, the uncertainty in our $L_\textmd{IR}$ measurement can be estimated from the variation in the SED shapes between individual galaxies. However, we cannot directly measure the SED shapes for our \bootes\ sample, because there is only one photometric band of far-IR observations available. 
Therefore, we took the 39 $z\sim1$ sources from K12, and selected a sub-sample of 25 SF-dominated (with mid-IR AGN fraction less than $10\%$, see K12 for the details in mid-IR spectral decomposition), far-IR detected (with at least two bands of SPIRE photometry) sources. This sub-sample spans a redshift range of $0.47<z<1.24$ and has at least 5 photometric data points from $24$ to $500\;\micron$; thus it can be used to estimate the dispersion of $S_{250}/L_\textmd{IR}$ due to the variation in SEDs for a population of star-forming galaxies. \par

We first estimated $L_\textmd{IR}$ and the shape of SED for each source in the K12 sub-sample by taking the available photometry and calculated the corresponding rest-frame monochromatic luminosity ($L_\lambda$). Combining $L_\lambda$ for each photometric bands, we calculated a best-fitting spline curve, then integrated along the spline curve in the rest-frame $20-300\;\micron$ range. 
For wavelengths beyond the longest wavelength of the spline curve, we used a linear interpolation by assuming that this part of the SED follows a Rayleigh-Jeans distribution. 
Since this sub-sample from K12 is selected to be SF-dominated and have far-IR constraints with at least two photometric bands, the best-fitting spline curves can trace the simple shapes of SF-related SEDs at this wavelength range. The integrated $20-300\;\micron$ luminosity ($L_{20-300}$) probes the thermal radiation from warm and cold dust, and represents the bulk of the total $8-1000\;\micron$ $L_\textmd{IR}$ (e.g. for the K12 template, $L_{20-300}\sim0.91L_\textmd{IR}$), thus we can use $L_{20-300}$ as a good proxy of $L_\textmd{IR}$. Along the best-fitting spline curves, we calculated the observed frame $250\;\micron$ fluxes at the redshift range of our \bootes\ sample, $0.25<z<0.8$, which corresponds to rest-frame wavelengths from $200$ to $140\;\micron$. The $S_{250}/L_{20-300}$ for each source in the K12 sub-sample can therefore be estimated as a function of redshift. For the K12 composite SED, we also calculated the $S_{250}/L_{20-300}$ in $0.25<z<0.8$.  We compared the $S_{250}/L_{20-300}$ for each source to that of the K12 template at $0.25<z<0.8$, and found that the standard deviations in the differences of $S_{250}/L_{20-300}$ between the K12 sources and the K12 template is $\sim 0.17$ dex at $z=0.25$ and $\sim 0.10$ dex at $z=0.8$. This shows that the deviations in far-IR spectral shapes across a representative population of star-forming galaxies are reasonably small, thus we can confidently estimate the star formation related $L_\textmd{IR}$ using Eq.~\ref{eq:lir} from the observed monochromatic $250\;\micron$ flux.  Even though the dispersion is lower at higher redshift, we conservatively chose 0.17 dex as the uncertainty in our $L_\textmd{IR}$ estimation. This is also consistent with the 0.17 dex uncertainty inferred by K12 for the $L_\textmd{IR}^\textmd{T}$ of the composite template. \par

As a check, we compared the $L_\textmd{IR}$ for star-forming galaxies from our method ($L_\textmd{IR}^{250}$) with the $L_\textmd{IR}$ obtained by directly fitting the {\it Spitzer} MIPS $24\;\micron$ flux to the  CE01 library ($L_\textmd{IR}^{24}$), which has been shown to be able to robustly estimate $L_\textmd{IR}$ for star-forming galaxies out to $z\sim 1$ \citep{magn09},
and found that the average difference between $L_\textmd{IR}^{250}$ and $L_\textmd{IR}^{24}$ is $\sim 0.1$ dex. However, we note that the difference is much larger for objects with significant AGN contamination (and thus bluer $S_{250}/S_{24}$), highlighting the need for far-IR data to measure SFR in AGN hosts.\par

\begin{figure}
\epsscale{1.05}
\plotone{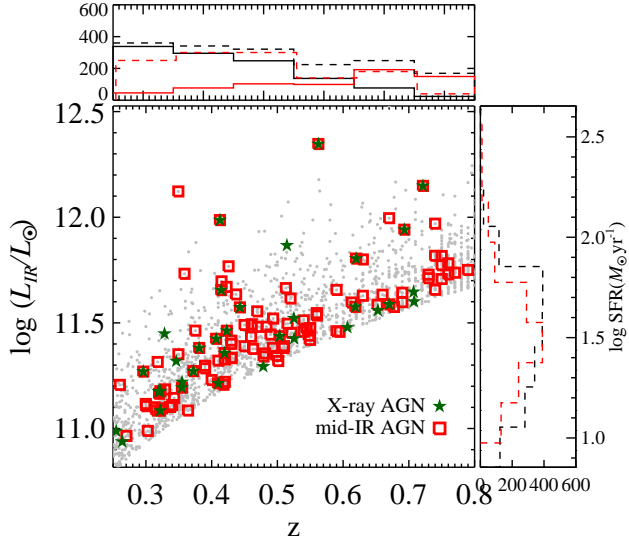}
\caption{The distribution of redshifts and $8-1000 \micron$ IR luminosities for our sample galaxies detected by both {\it Herschel} SPIRE $250\;\micron$ and {\it Spitzer} MIPS $24\;\micron$. X-ray AGNs are marked as green stars and mid-IR AGNs are marked as red squares. The histograms in redshift and $L_\textmd{IR}$ are also shown in the top and right panels. In the top panel, we show the different redshift distributions of the sources with only photometric redshifts (solid red line) and the sources with only spectroscopic redshifts (solid black line). The redshift distributions of AGNs (red dashed line, normalized to scale) and star-forming galaxies are also shown. In the right panel, we show that AGNs (red dashed line) and star-forming galaxies (black dashed line) have similar distributions in $L_\textmd{IR}$ (the histogram of AGNs is normalized to scale). These distributions show that the galaxies with identified AGN in our sample have distributions in redshift and $L_\textmd{IR}$ similar to those of star-forming galaxies.}
\label{fig:lir}
\end{figure}

The star formation rates for our sample were derived from $L_\textmd{IR}$ using the relation from \cite{kenn98araa}, modified for a Chabrier IMF \citep{chab03IMF,sali07sfr}:
\begin{equation} \label{eq:SFR}
\frac{\textmd{SFR}}{M_{\odot}\;\textmd{yr}^{-1}}=1.09\times10^{-10} \left(\frac{L_\textmd{IR}}{L_\odot}\right).
\end{equation} 
Fig.~\ref{fig:lir} shows the distribution in redshift and $L_\textmd{IR}$ of our sample. The comparison between the photometric and spectroscopic samples, and the comparison between the $L_\textmd{IR}$ of AGNs and star-forming galaxies are also shown as the normalized histograms on the side panels. We note that AGNs and star-forming galaxies have similar distributions in $L_\textmd{IR}$ and redshift, suggesting that there is no apparent difference in star formation properties between the AGN host galaxies and star-forming galaxies in our sample. 

\subsection{Black Hole Accretion Rate}
The large observed nuclear luminosities in AGNs are direct manifestations of black holes growth through mass accretion \citep[see, e.g.][]{bhreview12}. In this work, we calculated the rest-frame 2-10 keV X-ray luminosity ($L_\textmd{X}$) for direct comparison with other studies. The {\it k}-corrections were calculated based on an X-ray spectral index of 1.7 and a galactic gas column density of $N_H\sim10^{20}$ cm$^{-2}$. We chose $L_\textmd{X}$ as a proxy to estimate the black hole accretion rate: 

\begin{equation} \label{eq:BHAR}
\dot{M}_\textmd{BH}=0.15\frac{\epsilon}{0.1}\frac{22.4L_\textmd{X}}{10^{45}\textmd{ergs}^{-1}}M_{\odot}\;\textmd{ yr}^{-1}.
\end{equation}
For simplicity, $\dot{M}_{BH}$ in Eq.~\ref{eq:BHAR} is derived from $L_\textmd{X}$ with a constant bolometric correction factor of 22.4 \citep[the mean bolometric correction factor from][which is based on a sample of local, $L_\textmd{X}$=$10^{41-46}$ erg s$^{-1}$ AGN.]{vasu07bolc} Here $\epsilon$ is the mass-energy conversion efficiency \citep[we used a typical value $\epsilon\sim 0.1$, see][]{marc04smbh}. In the following paragraphs, we describe the methods we used to calculate the average $L_\textmd{X}$ for all of the star-forming galaxies in our sample. 

\subsubsection{Mid-IR AGN}
Since some of the actively accreting AGNs are not detected in X-rays due to obscuration, it is important to take them into account when calculating the average SMBH accretion rate for a population of galaxies. In particular, the X-ray stacking analysis in \cite{hickox09} shows that mid-IR AGNs without direct X-ray detections have higher average X-ray luminosity and harder X-ray spectra than those of star-forming galaxies at similar redshifts. We therefore assumed that the mid-IR AGNs without X-ray detections are faint in observed X-rays because of obscuration. 
Since both $L_\textmd{X}$ and the rest-frame $4.5\;\micron$ flux density ($L_{4.5}$) can be used to derive the bolometric AGN luminosity with a choice of bolometric correction factors, we first estimated $L_{4.5}$ for the mid-IR AGNs by interpolating the fluxes detected in all four IRAC bands. 
We next derived an empirical relation between $L_{4.5}$ and $L_\textmd{X}$ for all of the 163 AGNs that are identified by IRAC color-color cuts and are also X-ray detected in the redshift range of $0.25<z<0.8$. 
We found that the median $L_{4.5}/L_\textmd{X}$ for these AGNs is 4.59, or, $L_{4.5}=4.59L_\textmd{X}$. From the distribution of $\log (L_{4.5}/L_\textmd{X})$, we also derived an uncertainty of $\sim 0.37$ dex in this ratio by fitting a normal distribution to it. We then estimated the $L_\textmd{X}$ from $L_{4.5}$ for the 74 mid-IR AGNs without X-ray detection in our main sample using this empirically derived relation. For convenience, we denote the X-ray luminosity derived from this empirical relation as $L_\textmd{X}^\textmd{IRAGN}$.

\subsubsection{X-ray Stacking of Star-forming Galaxies}
Our goal is to study the average BHAR in star-forming galaxies over a range of SFR. We derived the SFR using Eq.~\ref{eq:SFR} and divided the galaxies in our sample into bins of SFR with approximately equal size and the number of galaxies in each bin being at least $5 \sigma$ above the Poisson noise. To estimate the average X-ray luminosity for all of the star-forming galaxies, we used an X-ray stacking analysis to account for sources not individually detected in X-rays. We defined the stacked X-ray counts as the average number of background-subtracted photons detected within the $90\%$ point-spread function (PSF) energy encircled radius at 1.5 keV, $r_{90}$, where $r_{90}=1''+10''(\theta/10')^2$. Here $\theta$ is the off-axis angle from the {\it Chandra} optical axis\footnote{{\it Chandra} Proposers's Observatory Guide (POG), available at \url{http://cxc.harvard.edu/proposer/POG}.}. We adopted background surface brightnesses of 3.0 and 5.0 counts s$^{-1}$ deg$^{-2}$ for the 0.5-2 keV and 2-7 keV bands, based on the estimates of the diffuse background \citep{hick07obsagn}. We converted count rates (counts s$^{-1}$) to flux (ergs cm$^{-2}$ s$^{-1}$) using the conversion factors $6.0\times10^{-12}$ ergs cm$^{-2}$ count$^{-1}$ in the 0.5-2 keV band and $1.9\times10^{-11}$ ergs cm$^{-2}$ count$^{-1}$ in the 2-7 keV band.  The error in the flux can be directly estimated from the error in count rates, which can be calculated using an approximation: $\sigma_X=\sqrt{X+0.75}+1$, where $X$ is the number of counts \citep{gehr86}. To estimate the average X-ray stacking luminosity from the X-ray flux, we assumed that all galaxies in each bin of SFR reside at the average luminosity distance for the galaxies in that bin. The uncertainty of the stacked X-ray luminosity can be derived from the combination of errors in the flux and the average luminosity distance. More details of the stacking procedure are described in \S 5.1 of \cite{hick07obsagn}. \par

\subsubsection{Average X-ray Luminosity}
Combining the contributions of X-ray and mid-IR identified AGNs as well as undetected sources, the average $L_\textmd{X}$ for star-forming galaxies can now be calculated in each bin of SFR: 
\begin{equation} \label{eq:avglx}
\begin{aligned} [c]
&\avg{L_\textmd{X}}=\\
&\left[\sum_{i=1}^{N_\textmd{XAGN}} (L_\textmd{X}^\textmd{XAGN})_i+\sum_{i=1}^{N_\textmd{IRAGN}}(L_\textmd{X}^\textmd{IRAGN})_i+N_\textmd{SFG} L_\textmd{X}^\textmd{stacking} \right]  \\
&\left[N_\textmd{XAGN}+N_\textmd{IRAGN}+N_\textmd{SFG}\right]^{-1}. 
\end{aligned}
\end{equation}

Here $N_\textmd{XAGN}$, $N_\textmd{IRAGN}$ and $N_\textmd{SFG}$ are the total numbers of X-ray identified AGNs, mid-IR identified AGNs without direct X-ray detections and star-forming galaxies without identified AGN, respectively. $L_\textmd{X}^\textmd{XAGN}$ and $L_\textmd{X}^\textmd{IRAGN}$ are the X-ray luminosities for individual X-ray AGN and the mid-IR AGN, and $L_\textmd{X}^\textmd{stacking}$ is the average X-ray luminosity from the stacking analysis for all star-forming galaxies without direct AGN detections in each bin. \par
We estimated our errors by propagating the observed uncertainties for $L_\textmd{X}^\textmd{XAGN}$, and the uncertainties for $L_\textmd{X}^\textmd{IRAGN}$ and $L_\textmd{X}^\textmd{stacking}$ estimated in \S2.6.1 and \S2.6.2 with a bootstrap method. The uncertainties in $L_\textmd{IR}$ were also taken into our bootstrap analysis. In each bootstrapping subsample, we first randomly resampled our sources with replacements, then replaced the original $L_\textmd{IR}$ for each source using a random normal error with an $1\sigma$ value of 0.17 dex. We then re-binned the random sample using the same bins. For each bin, we recalculated the stacked ${L_\textmd{X}}^\textmd{stacking}$ and the uncertainties for the sources that were not identified as AGNs, then replaced the $L_\textmd{X}$ for every detected AGN in the bin with a new $L_\textmd{X}$ within the normal error of the original AGN $L_\textmd{X}$. Finally, we recalculated the $\avg{L_\textmd{X}}$ and the average $L_\textmd{IR}$ for each bin. We repeated the bootstrapping 5,000 times, at which the variances in $\avg{L_\textmd{X}}$ and $L_\textmd{IR}$ converge to finite values. The results are shown in Fig.~\ref{fig:avg}, Fig.~\ref{fig:sup} and are discussed in the next section.\par

\section{Results}
In this section, we discuss the correlation between the average BHAR and SFR in star forming galaxies. We divided the galaxies in our sample into bins of SFR, and calculated the {\em average} black hole accretion rate in each bin, yielding an approximately linear correlation between the $L_\textmd{IR}$ and the average X-ray luminosity.

\subsection{The SFR-BHAR Correlation}
Using Eq.~\ref{eq:avglx}, we can calculate the average $L_\textmd{X}$ in each bin of SFR. However, it is well known that high mass and low mass X-ray binaries (HMXBs and LMXBs) can also generate X-ray luminosity that is correlated with SFR \citep[e.g.][]{grim03,rana03sf,gilf04,lehm10sfx}. To accurately estimate the SMBH accretion rate, we calculated the X-ray luminosities related to star-forming processes in each bin of SFR using the equation $L_\textmd{X}^\textmd{SF} = \alpha M_\star+\beta SFR$, which is the $\textmd{SFR}-L_\textmd{X}$ relation for HMXBs and LMXBs in \cite{lehm10sfx}. In this equation, the stellar mass $M_\star$ is only weakly correlated with SFR in active star-forming galaxies (e.g. SFR$>5M_\sun$ yr$^-1$). Thus for our sample and the Chabrier IMF we adopted, the equation can be re-written into $L_\textmd{X}^\textmd{SF} = 10^{26.4} \textmd{SFR}^{0.3}+10^{39.3}\textmd{SFR}$ \citep[see Eq. 3 in][for more details]{syme11lirlx}. \par

In addition, we also tested whether the limited volume of the sample could affect the BHAR-SFR correlation. AGNs with the highest luminosity are rare in this redshift range (and so might not be detected in our survey volume) but may contribute significantly to the $\avg{L_\textmd{X}}$ of our sample. 
We estimated the contribution of these rare, extremely luminous AGNs to the $\avg{L_\textmd{X}}$ using the X-ray luminosity function (XLF) from \cite{aird10xlf}. We note that the AGN XLF was not designed to represent the X-ray luminosity from the sources without direct X-ray observations, thus we first estimated the effect of limited volume on the sources that were identified as AGNs in our sample, i.e. $L_\textmd{X}>10^{42}$ erg s$^{-1}$; then we calculated the effect on the full population by adjusting the result from detected AGNs based on the AGN detection fraction. In detail, at the average redshift range of each SFR bin of our sample, we first calculated the ``intrinsic'' average AGN luminosity by directly integrating the XLF at $L_\textmd{X}>10^{42}$ erg s$^{-1}$. Then we estimated the ``detected'' average AGN luminosity by integrating the XLF with a high-end cutoff luminosity, at which the number of the detected AGNs in the volume of each SFR bin is $\leq 1$. 
We found that the difference between the ``intrinsic'' average AGN luminosity and the ``detected'' average AGN luminosity  is $10\%$ to $3\%$ from the first bin to the last bin of SFR in our sample. After the adjustments of the AGN detection fraction in each bin, the corrections on $\avg{L_\textmd{X}}$ would become $6.4\%,\;2.5\%,\;2.2\%$ and $2.9\%$, respectively. These corrections are small and do not make notable difference to our study of BHAR-SFR correlation in the large volume of the \bootes\ survey region, but might be important when calculating $\avg{L_\textmd{X}}$ for a sample with smaller volume. To accurately describe the correlation between the average SMBH growth and star formation, we subtracted our $\avg{L_\textmd{X}}$ with $L_\textmd{X}^\textmd{SF}$, and also increased our observed $\avg{L_\textmd{X}}$ values to account for volume effects as described above. The BHAR were then derived using Eq.~\ref{eq:BHAR}.\par
 
\begin{figure}[!t]
\epsscale{1.2}
\plotone{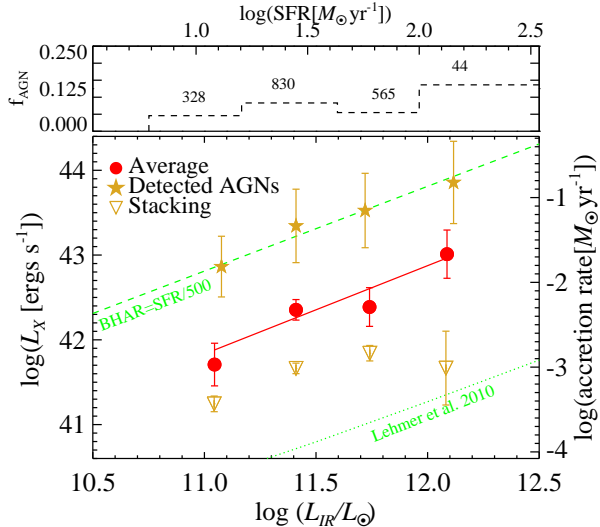}
\caption{The SFR-BHAR relation, calculated from the 8-1000 $\micron$ $L_\textmd{IR}$ and the 2-10 keV $L_\textmd{X}$ with Eq.s (2) and (3) for the entire $250\;\micron$ sample of 1,767 galaxies (red circles). The X-ray stacking luminosity for the X-ray non-detections are shown as downward triangles and the X-ray/IRAC selected AGNs are shown as stars. The sources are binned in approximately equal size SFR bins, with the vertical bars showing the errors from bootstrap re-sampling in each bin. The data points are plotted on the average $L_\textmd{IR}$ of each bin. The dashed green line on the top is $\dot{M}_{BH}=SFR/500$, the dotted green line in the bottom is the \cite{lehm10sfx} SFR-$L_\textmd{X}$ relation. In the top panel, we also present the AGN detection fraction ($f_{AGN}$) with the total number of galaxies in each bin. The width of each bin in the histogram covers the SFR range in the bin. This figure shows that the {\it average} BHAR is strongly correlated with the SFR in all rapidly star-forming galaxies. }
\label{fig:avg}
\end{figure}

We have determined that the average BHAR has a correlation to the SFR in our sample. The results are shown in Fig.~\ref{fig:avg}. The average X-ray luminosities, $\avg{L_\textmd{X}}$ as determined in Eq. 4 are shown as the red circles. The observed $L_\textmd{X}$ for the AGNs identified through X-ray or IRAC observations are shown as the stars, while the stacked $L_\textmd{X}$ for star-forming galaxies without direct X-ray observations are shown as the downward triangles. For comparison, we present the SFR-$L_\textmd{X}$ relation from the \cite{lehm10sfx} in  Fig.~\ref{fig:avg}. 
We also show the SFR-BHAR correlation corresponding to the local ratio of $M_\textmd{BH}$ and $M_\textmd{bulge}$ as the green, dashed line on the top of Fig.~\ref{fig:avg}. This $\dot{M}_{BH}=SFR/500$ relation is directly derived from the $M_\textmd{BH}/M_\textmd{Bulge}$ ratio found in \cite{marc04smbh}. Since the average $L_\textmd{X}$ of detected AGNs is subject to the flux limit in the observations, the fact that our data points for detected AGNs sit on the $\dot{M}_\textmd{BH}=\textmd{SFR}/500$ relation is only a coincidence. \par

The relation between $L_\textmd{IR}$ and $\avg{L_\textmd{X}}$ as shown in Fig.~\ref{fig:avg} can be fitted with a linear relation given by:
\begin{equation} \label{eq:avg}
\begin{aligned}[c]
&\log (L_\textmd{X} [\textmd{erg s}^{-1}]) =  \\
&(30.37\pm3.80)+(1.05\pm0.33)\log (L_\textmd{IR}/L_{\odot}), 
\end{aligned}
\end{equation}
which is calculated using the non-linear least squares fitting program MPFIT in IDL \citep{mpfit}. The uncertainties in the averages are estimated using bootstrap re-sampling in each bin. The reduced $\chi^2$ of this fitting is 0.99. An equation correlating SFR and BHAR can immediately be derived using Eq.~\ref{eq:SFR}, Eq.~\ref{eq:BHAR} and Eq.~\ref{eq:avg}: 

\begin{equation} \label{eq:main}
\begin{aligned}[c]
&(\textmd{BHAR}/M_{\odot}\;\textmd{ yr}^{-1})=  \\
&10^{(-3.72\pm0.51)}\left(\frac{\textmd{SFR}}{M_\odot\;\textmd{yr}^{-1}}\right)^{(1.05\pm0.33)}.
\end{aligned}
\end{equation}
 For an SFR of 100 $M_\odot\;$yr$^{-1}$, this corresponds to a ratio of BHAR to SFR of $\log$(BHAR/SFR)$ \sim -3.6 \pm 0.2$,  where the error is derived from calculating this ratio in each bootstrapping subsample when deriving Eq.~\ref{eq:avg}. \par

 We also calculated the same correlation using only the sample with spectroscopic redshifts. In this calculation, we adopted the sampling weight $w_i$ to account for the spectroscopic redshift completeness of the AGES sample. The sampling weight is the combination of the sparse sampling weight that accounts for the random target selection incompleteness, the target assignment weight that address the fiber-allocation selection, and the redshift weight which accounts for the unsuccessful redshift measurement. The details of the sampling weight can be found in \S3.1 of \cite{hickox09} and \cite{AGES11}. Using the same methods, the SFR-BHAR relation for the AGES galaxies with only spectroscopic redshifts can be written as:

\begin{equation}
\begin{aligned}[c]
&\log (L_\textmd{X} [\textmd{erg s}^{-1}]) =  \\
&(29.39\pm4.72)+(1.14\pm0.41)\log (L_\textmd{IR}/L_{\odot}).
\end{aligned}
\end{equation}

Here the power-law index is only higher by $\sim 0.1$ comparing to Eq.~\ref{eq:avg}, which is still consistent with the result from our main sample. Since photometric redshift measurements are not subject to the choices of sampling weights as the spectroscopic sample, we chose the result from Eq.~\ref{eq:avg} and Eq.~\ref{eq:main} as our primary conclusion in this work. \par

In Fig.~\ref{fig:avg}, we noticed that our stacked $L_\textmd{X}$ is at least $\sim 0.7$ dex higher than the \cite{lehm10sfx} $L_\textmd{X}^\textmd{SF}$ in the first three bins, suggesting that the $L_\textmd{X}$ contribution from star formation in our stacked X-ray luminosity is less than $\sim 20\%$ in these bins. For the bin with the highest SFR, the stacked X-ray luminosity is still higher than $L_\textmd{X}^\textmd{SF}$ by a factor of two. This shows that while the galaxies we stacked were not identified as AGNs neither using the common X-AGN selection criterion of $L_\textmd{X}>10^{42}$ erg s$^{-1}$ nor the IRAC color-color selection method, a significant fraction of the average $L_\textmd{X}$ for these star-forming galaxies arises from SMBH accretions. We stress that the $\avg{L_\textmd{X}}$ used in our primary analysis is the average $L_\textmd{X}$ due to SMBH accretion only, which was determined by subtracting the expected SF contribution.
We note that in an X-ray stacking study in the same \bootes\ survey region, \cite{wats09} concluded that the spectroscopically selected late-type galaxies have their X-ray luminosities dominated by HMXBs. However, the sample in \cite{wats09} are galaxies with lower star formation rates at lower redshifts. The stacked $L_\textmd{X}$ in the lowest SFR bin in our work is still in agreement with their sample at comparable SFR. Using our methods, deeper far-IR observations would be required to probe the more moderate SFR galaxies studied by \cite{wats09}. \par

\subsection{Effects of Flux Limit}
We identify that there might be observational bias in the BHAR to SFR correlation in our sample due the flux limits. The limited flux would cause galaxies with high SFR to be preferentially found at higher redshift. This bias translates into the different average redshift in each bin ($\avg{z}=0.31,0.46,0.67,0.68$ in the 4 bins of SFR, from low to high.) 
Studies of redshift evolution of SFR density \citep[e.g.][]{hopk06sfhist,rodi11irlf} and BHAR density \citep[e.g.][]{silv09sfagn,aird10xlf} suggest that the SFR and BHAR densities are higher at higher redshift (up to $z\sim 2$). Thus, when dividing galaxies into bins of SFR in a flux-limited sample, even if the $L_\textmd{X}$ and $L_\textmd{IR}$ are completely uncorrelated in individual galaxies, the redshift evolutions of the XLF and the infrared luminosity function (IRLF) would naturally yield a $L_\textmd{X}$-$L_\textmd{IR}$ relation. \par
To account for the selection bias due to this effect and to test whether the redshift coevolution of the SFR density and the BHAR density is the dominant factor driving the $L_\textmd{X}$-$L_\textmd{IR}$ correlation observed in this work, we examined the sample by directly computing the effect of the redshift evolution of the X-ray luminosity density \citep[XLD, e.g.][A10 hereafter]{aird10xlf}. In principle, if there is no BHAR-SFR correlation in individual galaxies, the observed difference of the $\avg{L_\textmd{X}}$ between the bins with the lowest and the highest SFR should be consistent with the pure redshift evolution of the XLD. We found that in our sample, the pure redshift evolution of the A10 XLD in the range of the average redshifts in our bins would translate into a difference in $\avg{L_\textmd{X}}$ of 0.47 dex. \par

To address this issue more carefully, we also created a ``mock'' catalog of galaxies, in which redshift distributions similar to our sample were generated. We generated IR luminosities for the galaxies in the mock catalog based on the IRLF from \cite{rodi11irlf}. 
Based on the different normalizations of this IRLF and the XLF from A10, we only assigned X-ray luminosities to a fraction of the mock galaxies according to the XLF from A10. For the rest of galaxies, we assumed their X-ray luminosities to be 0. Since the IR luminosity distribution and X-ray luminosity distribution were derived independently, there is no intrinsic correlation between $L_\textmd{X}$ and $L_\textmd{IR}$ in our mock catalog. To test the effects of the redshift evolution in XLF and the possible Malmquist bias, we took the flux limits in X-ray and far-IR of our sample and applied them to the mock catalog, then repeated the calculations described in \S3.1 to obtain $\avg{L_\textmd{X}}$ in bins of SFR. 
We found a weak correlation between $L_\textmd{IR}$ and $\avg{L_\textmd{X}}$ in our mock catalog, $\log L_\textmd{X}=37.45\pm1.93+ (0.30\pm0.17)\log L_\textmd{IR}$. The difference in $\avg{L_\textmd{X}}$ between the bins with the highest and the lowest SFR is $0.38$ dex, which is similar to the effect of pure XLD evolution we estimated in the previous paragraph. In Eq.~\ref{eq:avg}, we found that there is at least $1.31$ dex difference in the $\avg{L_\textmd{X}}$ of the bins with the lowest and the highest SFR in our sample, indicating that most, if not all, of our observed trend is due to the intrinsic correlation between SFR and BHAR. \par

\subsection{Comparison to Previous Studies}
\begin{figure}[!t]
\epsscale{1.3}
\plotone{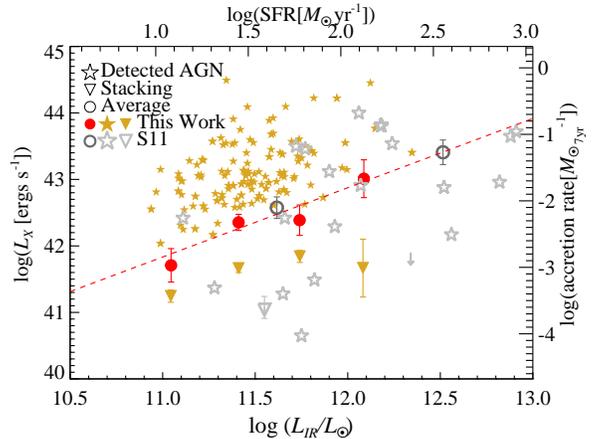}
\caption{Comparison of our result from Fig. 3 (the filled symbols) to the SPIRE-selected, $z\sim 1$ sample in the CDF-N field \citep[][the open symbols]{syme11lirlx}, which is covered by a much deeper, pencil-beam sized field. The axis, symbols and the error bars have the same meanings as in Fig.~\ref{fig:avg}. The correlation of Eq.~\ref{eq:avg} is shown as the dashed line. This comparison shows that even though the correlation between the $L_\textmd{X}$ and $L_\textmd{IR}$ for identified AGNs varies with the depth of the observations, the {\it average} correlation is consistent.}
\label{fig:sup}
\end{figure}

To examine whether the average SFR-BHAR correlation is subject to the limiting fluxes of the observations, we compared our result with the sample of {\it Herschel} selected star-forming galaxies in the pencil-beam Chandra Deep Field-North \citep{syme11lirlx} at redshift $z \sim 1$.  
From Table 2 in \cite{syme11lirlx}, we selected the galaxies with hard (2-10 keV) X-ray detections and $L_\textmd{IR}$ larger than $10^{11}L_\odot$, in which the average X-ray luminosity for the X-ray non-detected galaxies have been estimated.
For these LIRGs and ULIRGs, we calculated the average luminosity and the error in the 2-10 keV X-ray using a bootstrap resampling method similar to that we used for the \bootes\ data. Since there is no stacking signal in the ULIRG bin, we used the lower limit ($L_\textmd{X}=0$) for the X-ray non-detected sources in that bin to estimate the error in $\avg{L_\textmd{X}}$ conservatively. We also estimated the effects of the limited volume in this field using a similar approach described in the second paragraph of \S3.1, and found that the possible non-detections of the rare, high-luminosity AGNs might decrease the $\avg{L_\textmd{X}}$ by $\sim 26\%$ in the LIRG bin, and $\sim 7\%$ in the ULIRG bin. For the $\avg{L_\textmd{X}}$ in both our sample and the \cite{syme11lirlx} sample, we subtracted by $L_\textmd{X}^\textmd{SF}$ and made adjustments to account for the bias due to limited volume. A comparison of the results are displayed in Fig.~\ref{fig:sup},
which shows that in samples of star-forming galaxies with different X-ray flux limits, even though the average $L_\textmd{X}$ for the detected AGNs are different (so are the average $L_\textmd{X}$ for the galaxies without direct X-ray detections), 
the {\it average} $L_\textmd{X}$ to $L_\textmd{IR}$ relations are consistent. \par

\section{Discussion}

In the previous section, we presented the SFR to average BHAR correlation we found for the far-IR selected star-forming galaxies. 
From our X-ray stacking analysis, we found that the average $L_\textmd{X}$ of star-forming galaxies in our sample has and X-ray luminosity dominated by SMBH accretion instead of SF. This implies that black hole growth is not limited to the detected AGNs only.
Recent studies also argue for a picture that departs from a simple duty cycle scenario for SMBH accretion. In the duty cycle scenario, SMBHs accrete the most of their mass during short episodes of accretion near Eddington limit, and are relatively quiescent otherwise. In comparison, \cite{hopk09qso},\cite{aird12agnedd} and \cite{bong12} show that a substantial population of AGNs spend most of their lifetime accreting at lower Eddington ratio. Hence the SMBH growth during the lower Eddington ratio state cannot be neglected. In this vein, and considering that AGN may vary by over 5 orders of magnitude in Eddington ratio on a timescale much shorter than that of galactic star formation \citep[e.g.][]{hickox09,hopk09qso,nova11,aird12agnedd,bong12}, we argue that the {\it average} BHAR is a more appropriate tracer to study the correlation to the SFR, since the detected AGNs are only a small fraction of accreting SMBHs residing at the higher end of the Eddington ratio distribution. We have found evidence consistent with a universal BHAR-SFR correlation (Eq.~\ref{eq:avg} and Eq.~\ref{eq:main}). 
This result is consistent with a simple picture in which the BHAR-SFR link exists in star-forming galaxies over a wide range of SFR. We argue that the discrepancy between our result and the scenario where AGN and star formation are only linked in the most rapidly growing systems can be attributed to the timescale difference between the variability of AGN accretion efficiency and star formation. \citep[][Hickox et al., in preparation]{mull12agnms}. \par

We note that the observed ratio between BHAR and SFR in our work is lower than the observed black hole mass ($M_\textmd{BH}$) to galaxy bulge mass ($M_\textmd{Bulge}$) ratio for local galaxies. Other studies have also obtained the BHAR to SFR ratios consistent with the values observed here \citep{raff11agnsf_aph,mull12agnms}. Different arguments have been proposed to explain the low BHAR to SFR ratio. 
One is that the SF in disk galaxies tends to concentrate in their disks, which leads to a BHAR to galaxy-wide SFR ratio lower than that inferred by the observed $M_\textmd{BH}$ and $M_\textmd{Bulge}$ \citep[e.g.][]{jahn09,cist11,cist11agn,raff11agnsf_aph}. 
In addition, the average BHAR might be underestimated due to non-detections of a substantial population of heavily obscured AGNs, which can be responsible for as much as $50\%$ of SMBH growth \citep[e.g.][]{gill07cxb,trei09,mull12agnms}. 
\par

In summary, we studied the average BHAR for a sample of star-forming galaxies with SFR measurements without contamination from AGN using {\it Herschel}. We used AGNs selected at X-ray and mid-IR wavelengths to ensure that our BHAR is not biased by AGN obscuration, and employed an X-ray stacking analysis to measure SMBH accretion for star-forming galaxies without direct X-ray detections. We obtained an almost linear relation between the average BHAR and SFR of $\log \textmd{BHAR} =(-3.72\pm0.52)+(1.05\pm0.33)\log\textmd{SFR}$, and determined that this relation also holds for deeper, narrower  observations, suggesting that the average BHAR to SFR correlation is a universal consequence of the coevolution between SMBHs and galaxies. The next step of understanding the SFR to BHAR correlation in different populations of galaxies requires information on the {\it distribution} of AGN X-ray luminosity as a function of SFR, which will only be possible with a wide, deep X-ray survey. \par

\begin{acknowledgements}
We are grateful to the anonymous referee for the very careful reading of the paper, and the suggestions that significantly improved this paper. We thank our colleagues on the AGES, IRAC Shallow Survey, SDWFS, NDWFS and the \xbootes\ teams, and the HerMES team for making the data publicly available. The first {\em Spitzer} MIPS survey of the \bootes\ region was obtained using GTO time provided by the {\em Spitzer} Infrared Spectrograph Team (PI: James Houck) and by M. Rieke.  We thank the collaborators in that work for access to the $24\;\micron$ catalog generated from those data by Emeric LeFloc'h. We also thank Chris S. Kochanek for useful discussions. C.-T.J.C was supported by a Dartmouth Fellowship. This work was supported in part by Chandra grants SP8-9001X and AR8-9017X.
\end{acknowledgements}

\end{document}